\documentclass[12pt]{article}
\usepackage{amsmath}
\usepackage{graphicx}
\def\\{\hfill\break} \let\==\equiv

\let\0=\noindent

\def\qed{\hfill\raise1pt\hbox{\vrule height5pt width5pt depth0pt}}

\def\be{\begin{equation}}
\def\ee{\end{equation}}
\def\bea{\begin{eqnarray}}\def\eea{\end{eqnarray}}
\def\be{\begin{equation}}
\def\ee{\end{equation}}
\def\bea{\begin{eqnarray}}
\def\eea{\end{eqnarray}}

\def\hsp5{\hspace{5mm}}

\def\case#1/#2{\textstyle\frac{#1}{#2}}
\newcommand{\bc}{\begin{center}}
\newcommand{\ec}{\end{center}}
\newcommand{\ber}{\begin{eqnarray*}}
\newcommand{\ear}{\end{eqnarray*}}
\newcommand{\ba}{\begin{array}}
\newcommand{\ea}{\end{array}}

\newcommand{\hs}{\,-\,}

\newcommand{\ei}{\end{itemize}}

\newcommand{\bras}[1]{\left[#1\right]}
\newcommand{\brac}[1]{\left\{#1\right\}}

\newcommand{\nab}{\nabla}

\newcommand \veps {\varepsilon} 

\newcommand{\A}{{\cal A}}
\newcommand{\E}{{\cal E}}
\renewcommand{\H}{{\cal H}}

\def\case#1/#2{\textstyle\frac{#1}{#2} }

\title{Variations on Birkhoff's theorem}
\author{George F R Ellis,\\
 \emph{Mathematics Department and ACGC},\\
 \emph{University of Cape Town, Cape Town, 7700};\\ \\
Rituparno Goswami,\\
\emph{Astrophysics and Cosmology Research Unit},\\
 \emph{School of Mathematics Statistics and Computer Science},\\
\emph{University of KwaZulu-Natal, Durban, 4000}.}

\begin{document}
\maketitle 
\begin{abstract}
The relation between the expanding universe and local vacuum
solutions, such as that for the Solar System, is crucially mediated by
Birkhoff's theorem. Here we consider how that relation works, and
give generalizations of Birkhoff's theorem when there are geometric
and matter and perturbations. The issue of to what degree dark matter
might influence the solar system emerges as a significant question.
\end{abstract}
\section{Birkhoff's theorem}
Birkhoff's theorem in General Relativity
\cite{Bir23,HawEll73,Din92}, actually first discovered by Jebsen
\cite{Jeb21} (see \cite{JohRav06}),  states that any local
spherically symmetric solution of the vacuum Einstein field
equations
\begin{equation}\label{efe}
    R_{\mu\nu}   = 0
\end{equation}
must admit an extra Killing vector, and so be part of the
Schwarzschild vacuum metric:
  \begin{equation}\label{Sch}
ds^2 = -\left(1 - \frac{2m}{r}\right) + \frac{dr^2}{1 -
\frac{2m}{r}}+ r^2 \left(d\vartheta^2 + \sin^2 d\varphi^2\right),
\end{equation}
where $m$ is a constant representing the mass of the central object
(if there is no central mass so that $m=0$, this is just flat
spacetime). If the Killing vector is timelike, the spacetime is
locally static, and is asymptotically flat if it extends far enough;
the local solution is part of the exterior region ($r > 2m)$ of
(\ref{Sch}). If the Killing vector is spacelike, the spacetime is
locally spatially homogeneous: it is part of the interior black hole
region ($r < 2m$) of (\ref{Sch}), and runs
into a singularity if it extends far enough.\\

The basic idea is that a spherically spherical object should produce
a spherically spherical gravitational field; another mass elsewhere,
or anisotropic boundary conditions, would disturb the spherical
symmetry. So the solution represents an isolated spherical object;
the theorem then says the exterior spacetime is static. It should be
emphasized that this is a local result in the sense that it only
depends on the field in some open domain $U$ being vacuum and
spherically symmetric in that domain; then that domain is either
static or spatially homogeneous. The proof is local (see section \ref{BirkhoffII} below), and so does not
directly use boundary conditions at infinity, although those
conditions will indirectly affect the outcome because they do
influence whether the region $U$ is spherically symmetric or not.\\

A consequence is that no radial changes in a spherical star (whether
it is expanding, pulsating, collapsing, or whatever) affect its
external gravitational field: all spherically symmetric vacuum
gravitational fields are indistinguishable for $r
> R_s> 2m$ where $R_s(t)$ is the coordinate value at the surface of the
star. This means a spherically pulsating star cannot emit
gravitational waves, nor gravitationally radiate away its mass. \\

In summary, the Schwarzschild solution is the unique spherically
symmetric solution of the vacuum Einstein field equations
(\ref{efe}): a spherically symmetric gravitational field in empty
space outside a star must be static, with a metric given by
(\ref{Sch}) for $r>2m$. It represents the spacetime of the Solar
System, and all other spherically symmetric star systems, to very
good approximation, and so is key in much astrophysics and
astronomy. \\

If the cosmological constant is non-zero, the result essentially
remains true. The Kottler metric \cite{Per04}, found independently
by Kottler \cite{Kot18} and Weyl \cite{Wey19}, is the unique
spherically symmetric solution of the field equations with a
cosmological constant $\Lambda\neq 0$:
\begin{equation}\label{efe1}
    R_{\mu\nu} + \Lambda g_{\mu\nu} = 0
\end{equation}
has solution
\begin{equation}\label{Kottler}
ds^2 = -\left(1 - \frac{2m}{r}-\frac{\Lambda r^2}{3}\right) +
\frac{dr^2}{1 - \frac{2m}{r}-\frac{\Lambda r^2}{3}}+ r^2
\left(d\vartheta^2 + \sin^2 d\varphi^2\right).
\end{equation}
It is also known as the Schwarzschild--de Sitter metric for
$\Lambda>0$  and the Schwarzschild--anti-de Sitter metric for
$\Lambda<0$. If $\Lambda=0$ it is the Schwarzschild solution; if
$m=0$, it is just the de Sitter or anti-de Sitter metric. The global
structure is complex, depending on the
relation between $m$ and $\Lambda$ (see \cite{LakRoe77}).\\

Birkhoff's theorem can be generalized to some matter fields: any
spherically symmetric solution of the Einstein---Maxwell field
equations must be static in the exterior domain, with a
Reissner-Nordstr\"{o}m metric (\cite{Din92}, section
18.1).\footnote{This is of academic interest only, as charged stars
do not exist in reality. If they did, astronomy would be governed by
electric forces rather than gravity, because gravity is such a weak
force.} However it does not hold for matter such as baryons or a
perfect fluid: their spherically symmetric solutions have a dynamic
behavior \cite{LTB}. But systems such as the solar system are not
dominated by a perfect fluid or any other continuous matter: they
are basically empty space with isolated bodies embedded in the
vacuum. To a large extent the same is true of galaxies, mainly made
up of isolated stars, and clusters of galaxies, mainly made up of
isolated galaxies. With a caveat about dark matter (see below), most
of the universe is empty space. In any case even if dark matter is
present, all these systems are presently either static, or at least
stationary, to a very good approximation. \\

This raises a dilemma: if almost all local stellar systems are
static, how can they be put together to give an expanding universe?
Putting static domains together to make an expanding spacetime is
non-trivial. That is the topic of Section \ref{expand}.\\

Birkhoff's theorem underlies the crucial importance in astrophysics
of the Schwarzschild solution, as it means that the exterior metric
of any exactly spherical star must be given by the Schwarzschild
metric and this also underlies the uniqueness results for
non-rotating black holes. However it is an exact theorem that is
only valid for exact spherical symmetry; but no real star system is
exactly spherically symmetric (for example if they have planets). So
a key question is whether the result is approximately true for
approximately spherically symmetric vacuum solutions. In Section
\ref{sec:perturbations}, we prove an ``almost Birkhoff theorem'' for
this case \cite{GosEll11} where spherical symmetry is not exact.
Furthermore the Solar System is not exactly empty: it is pervaded by
very low density material. In Section \ref{sec:finitge infinty} we
prove a second ``almost Birkhoff theorem'' for that case too
\cite{GosEll12}. So the results carry over to astrophysically
realistic situations (such as the Solar System), which are not empty
and where spherical symmetry is not exact. \\

The key issue that
becomes clear through this study is the question of whether or not
dark matter pervading systems such as the solar system links local
systems to the cosmological expansion. If it does, then in principle
one can measure the Hubble constant
by solar system measurements.\\

This paper is an extension of a talk at the Spanish Relativity
Society meeting ERE2012 at Guimar\~{a}es by one of us (GFRE).

\section{Birkhoff's theorem and the expanding universe}\label{expand}
Birkhoff's theorem plays a key role as regards the relation of the
expanding universe and local vacuum solutions. The issue is, does
the Hubble expansion affect local physical systems such as the solar
system? Birkhoff's theorem says no, it does not if the system is
empty except for a spherically symmetric central object, so that
spacetime is locally spherically symmetric. The solution is then
locally static and hence completely decoupled from the global
expansion. Therefore you cannot measure the Hubble constant in the
solar
system: it has decoupled from the universal expansion.\\

But this applies almost everywhere. The universe is largely
comprised of empty space, with isolated compact objects scattered in
the void. Galaxy clusters are mainly empty space, as are galaxies
themselves, with immense empty regions separating the isolated stars
that are the bulk of matter in galaxies. Indeed most of the baryonic
matter in galaxies is concentrated in isolated stars surrounded by
empty space. Most of the universe is locally static (or perhaps
stationary, if the matter is rotating). Now it is true that this
pictures is complicated by the issue of how clustered dark matter
is, given that dark matter is the bulk of matter on a cosmological
scales; so maybe it is not empty locally after all.  We will return
to this issue later: for the moment we concentrate on cases where
this is a good approximation.

\subsection{Vacuum almost everywhere}
Consider a universe that is locally made of spherically symmetric
vacuum regions (such as the Solar System), which are static, because
of Birkhoff's theorem. They need to somehow be joined together to
give a globally expanding, approximately spatially homogeneous
spacetime. Thus locally rigid domains are joined to give a globally
expanding spacetime. How is it done? We will look successively at
perturbed FLRW models (Section \ref{Pertflrw}); Swiss Cheese models
(Section \ref{swisscheese}); Lindquist-Wheeler models (Section
\ref{Lindquist-Wheeler}); and the exact two-mass version of those
models (Section \ref{twomass}).

\subsection{Perturbed FLRW models}\label{Pertflrw}
The standard models for structure formation in cosmology are
perturbed Friedmann-Lema\^{\i}tre-Robertson-Walker (FLRW) models
with inhomogeneities imbedded, as represented in a linearised
solution of the Einstein Equations \cite{EllBru89,MukFelBra92}.
However they are fluid filled everywhere, and so do not represent
the situation sketched above. The FLRW regions expand and carry the
inhomogeneities representing structure such as galaxies with them;
these models do not represent local empty static domains, such as
the Solar System. They may perhaps be able to locally represent
virialised regions that have in effect opted out of the cosmic
expansion, but whether this can be done in a \emph{single}
coordinate system that represents \emph{all} such locally static
systems is not clear.\\

One may possibly be able to do it for a region where the Hubble
velocities are small near the origin of one coordinate system, if
one uses a static non-comoving frame for that domain (see equation
(2.1) in \cite{Bauetal10} for steps in this direction); but this
does not give the desired representation for local static systems
anywhere else. Thus one can perhaps do this for a particle $Q$ at
the origin of coordinates $x^i(Q)$, correctly representing its local
domain $U(Q)$ as exactly static; or else for a different distant
particle $P$, correctly representing its local domain $U(P)$ as
static, using a different set of coordinates $x^i(P)$ centered in
$P$. But these are different coordinate systems. The problem is you
can't do it simultaneously for both regions using one single
coordinate system, nor \emph{a fortiori} for $10^{11}$ separate
static domains at all local locations across the entire visible
universe that abut each other with no intervening fluid filled
domains. But that is what
is needed to represent the locally static nature of spacetime everywhere.\\

Overall these models do not show how local static domains fit
together to give expansion, because they rely on an all-pervading
cosmic fluid model for their dynamics, and that fluid everywhere
embodies the expansion of the universe. It is perhaps not impossible
to show how one can get local static domains in such models, but
remember we have to do that \emph{everywhere}: on the view put here,
the universe is globally made up of locally static vacuum domains
with isolated stars imbedded in them. That is what those coordinates
probably cannot represent.  These models rather represent the case
where relative velocities decrease towards zero at small scales, but
are never actually zero for any finite distance, because there is an
expanding cosmic fluid everywhere. It is likely one could measure
the Hubble constant in the Solar system in such models, if we take
them as extending down to that scale.

\subsection{Swiss cheese models} \label{swisscheese}
The Einstein-Strauss ``Swiss Cheese'' models
\cite{EinStr45,EinStr45a} do indeed properly represent local static
domains such as the solar system.  They are based on a FLRW
background model with vacuum regions (``vacuoles'') cut out, and
carefully matched with the fluid solutions at the boundaries between
the empty regions and the cosmological fluid. The vacuum domains can
have static star models imbedded at the centre, to give non-singular
vacuoles imbedded in a FLRW model. One can do this for vast numbers
of such vacuoles, considered as representing spacetime around either
stars or galaxies, and and can even do it in a
hierarchical way that includes both \cite{Rib92,MurDye04}.\\

There is no problem with expansion in this case: the connected FLRW
regions surrounding the vacuoles do indeed expand, and carry the
static vacuoles with them. One cannot measure $H_0$ in the Solar
System in these models, as the vacuoles are static. However these
models do not solve the issue posed here, because the background
FLRW model in them  embodies the expansion and carries the vacuoles
apart from each other. We want to model the case where there are no
such fluid-filled domains occupying the space between stars or
galaxies.\\

Thus there is no problem with global expansion if globally
interconnected fluid domains are allowed to surround the static
vacua as in the Swiss Cheese models, but we want to know what
happens if there is vacuum everywhere except in interiors of stars.

\subsection{Lindquist-Wheeler}\label{Lindquist-Wheeler}
In an innovative paper, Lindquist and Wheeler \cite{LinWhe57}
accurately modelled the situation considered here by considering a
regular lattice of Schwarzschild vacuum cells joined together to
give an expanding solution. There are no fluid filled regions in
these models; it is entirely made of static domains with imbedded
stars, as envisaged above. They average to a FLRW spacetime with
closed spatial sections when considered on large scales, and an
effective Friedmann equation
results from the junction conditions at the boundaries between the static domains.\\

Thus unlike the previous two cases, there is NO background FLRW
spacetime in this case, and no connected fluid that expands and
carries the stars and galaxies with them. The FLRW model emerges as
a large scale averaged approximation \cite{LinWhe57}. The Locally
Static vacuum domains recede from each other because of the boundary
conditions at the joins between them. Ferreira and Clifton have
recently extended these models in interesting ways
\cite{CliFer09,Cli11}, using perturbation methods.\\

These are very good models in terms of tackling the issue posed
here, but they are not exact solutions: the junction conditions in
\cite{LinWhe57} are not very clear, because the lattice symmetry
means the spherical symmetry of the vacuoles is not exact. The
Schwarzschild solutions in each vacuole are an approximation to the
real geometry because of this anisotropy (although one can get exact
solutions for the corresponding initial value problem
\cite{CliRosTav12}). Clifton and Ferreira give helpful perturbative
solutions of this kind \cite{CliFer09,Cli11}; but the issue we want
to look at is, are there any exact solutions of this nature?

\subsection{Two mass solution}\label{twomass}
One can indeed find exact such solutions in a very simple case: a
2-mass solution of this kind, with compact spatial sections because
of a positive $\Lambda$ term \cite{UzaEllLar11}. In order to
understand how locally static configurations around gravitationally
bound bodies can be embedded in an expanding universe, that paper
investigate the general relativity solutions describing a space-time
whose spatial sections have the topology of a 3-sphere with two
identical masses at the poles. Introduction of a
cosmological constant allows closed space sections.\\

\begin{figure}[h!!]
\begin{center}
\includegraphics[height=5in]{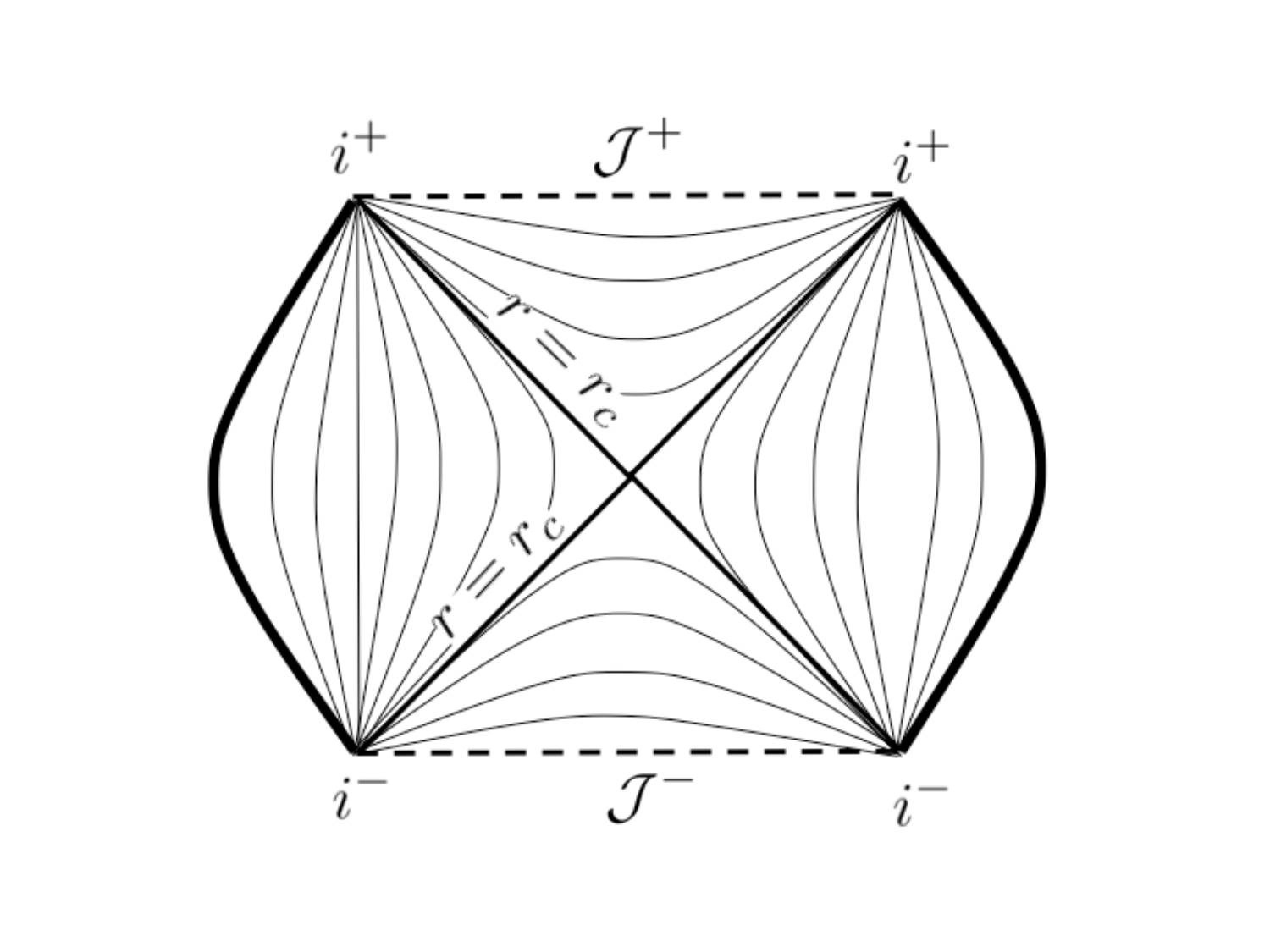}\label{join1}
\caption{\emph{The Penrose diagram for the two-mass Kottler solution
\cite{UzaEllLar11}.  There are no event horizons surrounding the
antipodal origins (these are filled in by interior star
solutions).}}
\end{center}
\end{figure}

One envisages two massive objects $M_1$, $M_2$ of equal mass $M$,
each embedded in a vacuum spherically symmetric domain $U_1$, $U_2$
to give a spherical stellar model surrounded by empty space (no
horizon occurs, because their surfaces $R_1$, $R_2$ lie outside
their Schwarzschild radii). Each vacuum domain is a segment of a
Kottler solution (\ref{Kottler}) with 2-sphere surface area at
coordinate value $r$ given by $A = 4 \pi r^2$. The cosmological
constant $\Lambda
> 0$ is chosen large enough so the area $A$ reaches a maximum at a
value $A_* = 4 \pi r^2$, and $U_1$, $U_2$ are joined back to back at
this value $r_*(\Lambda,M)$, to give closed space sections. Thus we
form a vacuum spacetime with compact spatial sections and identical
antipodal stellar
masses; it is spherically symmetric about each of the two masses.\\

The technical issue is matching the solutions at $r_*$ using the
Israel junction conditions: the first and second fundamental forms
must be continuous so that  no surface layer occurs. Because $r_*$
is a constant (it can depend only on $M$ and $\Lambda$, which are
both constant) it seems at first glance that  this model cannot
expand: only solutions with a fixed size are possible. One should
obtain the analogue of the Einstein Static universe:  a two mass
exact solution that is static, and vacuum everywhere except at the
two antipodal stars. \\

\begin{figure}[h!!]
\begin{center}\label{join2}
\includegraphics[height=5in]{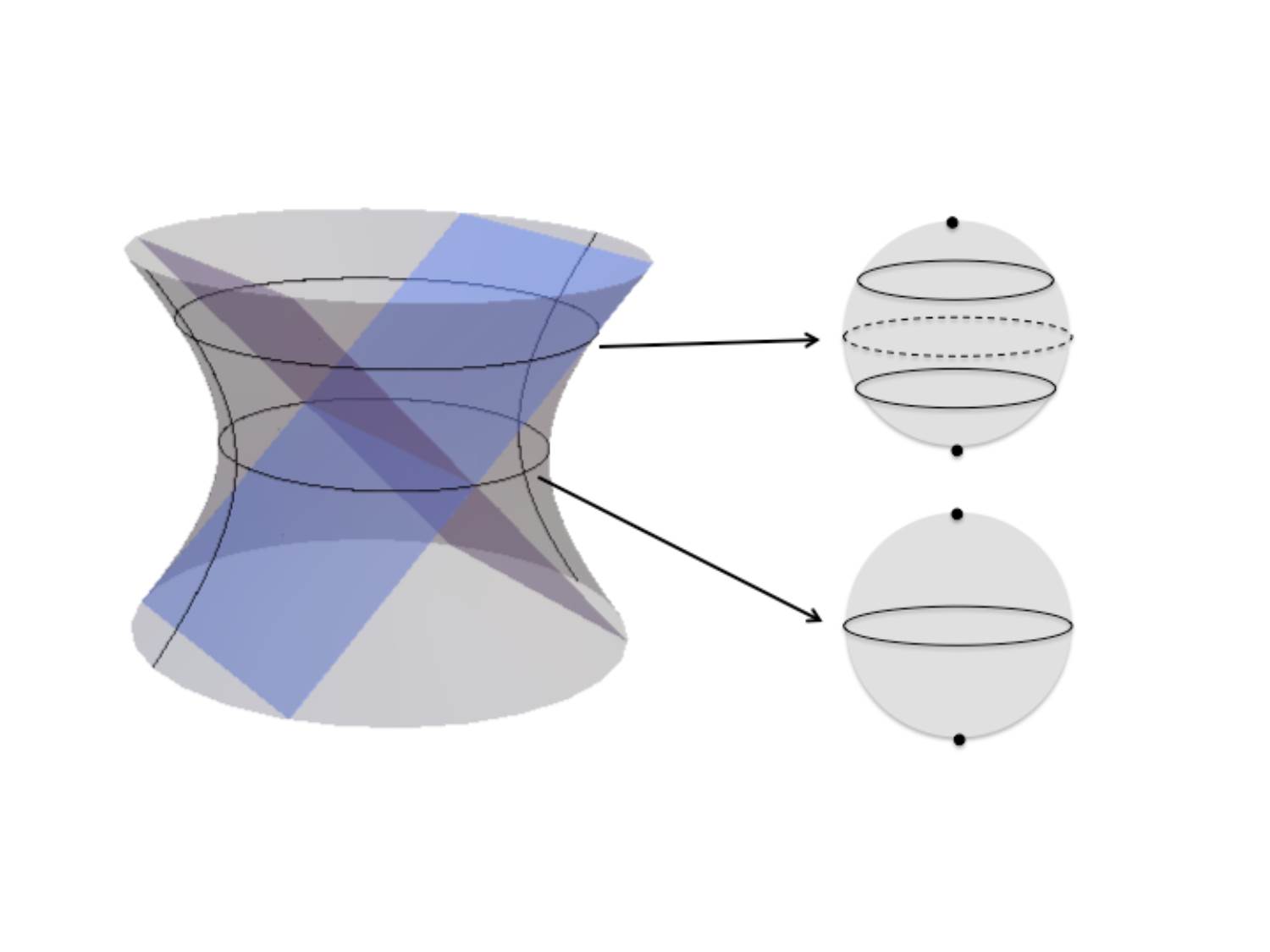}
\caption{\emph{Embedding of the 2-mass solution in 5-d spacetime: De
Sitter hyperboloid, with two antipodal masses imbedded
\cite{UzaEllLar11}. Inset: the spatial sections $\{t = const.\}$ at
a generic time (above) and at the time symmetric throat (below).}}
\end{center}
\end{figure}

However closer inspection shows that such static solutions are not
possible: one can't simultaneously match both the spatial and time
components of the second fundamental form. Israel junction
conditions imply that two spherically symmetric static regions
around the masses cannot be glued together in the desired way. The
resolution \cite{UzaEllLar11} is that one must match across a null
horizon, like the way de Sitter universe has local static domains
matched to expanding domains across null horizons
\cite{Sch59,HawEll73}. The Penrose diagram for this situation is
shown in Figure 1, and an embedding diagram in Figure 2.\\

The de Sitter static frame covers only part of the de Sitter
hyperboloid; taking spatial sections $\{t = constant\}$ in the
imbedding diagram, those static sections are separated from each
other by expanding domains across 2-spheres that are the null
horizons (top inset in Figure 2); those spatial sections decrease
from infinity, reach a time-symmetric minimum (bottom inset in
Figure 2), and re-expand. The static model we tried to find is just
the situation at that minimum radius: the matching we sought is
possible at that instant (and only then), because the second
fundamental form instantaneously vanishes there. Thus study of the
extension of the Kottler space-time shows that there exists a
non-static solution consisting of two static regions surrounding the
masses, that are each matched to a Kantowski-Sachs expanding region
at a null horizon. It is the expanding vacuum domains that interpose
between the static domains that allows the universe to expand. \\

There must be a set of coordinates for this solution corresponding
to the de Sitter $k=+1$ frame with expansion parameter $a(t) = \cosh
Ht$. \textbf{To be done}:  find a global perturbed FLRW coordinate
system for this model, like a perturbed de Sitter hyperboloid, for
the case when the mass is very small; and see how to extend it to
cases where the mass is large.

\subsection{Dark matter and local systems}\label{darkmatter}
The two mass solution shows how static vacuum domains can be joined
together to form an expanding universe: you glue them together
across expanding vacuum regions. \\

Now if that the vacuum dominated
model proposed above is all wrong and there are no such vacuum
domains (e.g. in the solar system) because of dark matter, that
analysis will not apply: the universe can be built up out of
solar-system like domains because they are pervaded by dark matter
which takes part in the cosmic expansion, hence they are not exactly
static (as must be the case if they are vacuum filled domains,
because of Birkhoff's theorem). And in that case you will indeed be
able to measure the Hubble constant in the solar system. The same
issue arises as regards the galaxy: if its internal dynamics is
dominated by dark matter, the question arises as to whether
that dark matter is taking part in the Hubble expansion or not. Thus there are two options:\\

\textbf{The locally static case}: The local domains may still be
essentially static, and something like the two mass model must
apply, even when there is not a vacuum. Indeed one may surmise this
is this indeed the case: galaxies are not in violent internal radial
motion, and the Hubble parameter does not enter analyses of galactic
dynamics \cite{BinTre94}. In that case the essential ideas of the
two mass solution will still apply, even thought the details will be
different: locally static regions can expand
because they are separated from each other by expanding domains. \\

\textbf{The locally non static case}: If this is not true, and a
very low density of dark matter pervading the solar system is indeed
taking part in the cosmic expansion, then these non-static domains
can be stitched together to give an expanding universe; and we can
least in principle measure the Hubble constant in the solar system,
opening up a new avenue of investigation in cosmology.

\section{Almost Birkhoff: perturbations}\label{sec:perturbations} %
The core content of Birkhoff's theorem \cite{Bir23} is that any
spherically symmetric solution of the vacuum field equations has an
extra symmetry: it must be either locally static or spatially
homogeneous. This underlies the crucial importance in astrophysics
of the Schwarzschild solution, as it means that the exterior metric
of any exactly spherical star must be given by the Schwarzschild
metric and it  also underlies the uniqueness results for
non-rotating black holes. However it is an exact theorem that is
only valid for exact spherical symmetry; but no real star is exactly
spherically symmetric. So a key question is whether the result is
approximately true for approximately spherically symmetric vacuum
solutions. In this section, we summarise the proof of an ``almost Birkhoff theorem''
for that case \cite{GosEll11}, so those results carry over to
astrophysically realistic situations (such as the Solar System)
where spherical symmetry is not exact.\\

There are of course many papers discussing perturbations of the
Schwarzschild solution, but none appear to focus on this specific
issue. The rigidity embodied in this property of the Einstein Field
Equations is specific to vacuum General Relativity solutions, or
solutions with a trace-free matter tensor. One should also note that
the result is local: both Birkhoff's theorem, and our generalization
of it, are independent of boundary conditions at infinity: they hold
in local neighborhoods $U$.

\subsection{1+1+2 covariant splitting of spacetime}
We prove the result by using the 1+1+2 covariant formalism
\cite{extension,CLa07}, which developed from the 1+3 covariant
formalism.\\

In the 1+3 covariant formalism \cite{Covariant,Ellis71}, first we
define a preferred timelike congruence with a timelike unit tangent
vector $u^a$ ($u^a u_a = -1$). Then the tangent spaces to spacetime
are split in the form $R\otimes V$ where $R$ denotes the timeline
along $u^a$ and $V$ is the tangent 3-space perpendicular to $u^a$.\\

In the (1+1+2) approach we further split the tangent 3-space $V$, by
introducing a spacelike unit vector $ e^{a} $ orthogonal to $ u^{a}
$ so that $ e_{a} u^{a} = 0\;,\; e_{a} e^{a} = 1$. Then the
\textit{projection tensor} $N_{a}{}^{b} \equiv g_{a}{}^{b} +
u_{a}u^{b} - e_{a}e^{b}$ projects vectors onto the tangent
2-surfaces orthogonal to $e^{a}$ and $u^a$, which, following
\cite{extension}, we will refer to as `{\it sheets}'. In the (1+3)
approach any second rank symmetric 4-tensor can be split into a
scalar along $u^a$, a 3-vector orthogonal to $u^a$, and a {\it
projected symmetric trace free} (PSTF) 3-tensor. In (1+1+2) slicing,
we take this split further by splitting the 3-vector and PSTF
3-tensors with respect to $e^a$. Any 3-vector, $\psi^{a}$, can be
irreducibly split into a component along $e^{a}$ and a sheet
component $\Psi^{a}$ orthogonal to $e^{a}$ i.e. $\psi^{a} = \Psi
e^{a} + \Psi^{a}$, where we have $\Psi\equiv\psi^{a} e_{a}$ and
$\Psi^{a} \equiv N^{ab}\psi_{b}$.  A similar decomposition can be
done for PSTF 3-tensor, $\psi_{ab}$, which can be split into scalar
(along $e^a$), 2-vector and PSTF 2-tensor.\\

The key variables of the 1+1+2 formalism obtained in this way are
(see ~\cite{GosEll11} for a detailed physical description of these
variables) \bea \left[ \Theta, \A, \Omega,\Sigma, \E, \H, \phi,\xi,
\A^{a},\Omega^{a}, \Sigma^{a}, \alpha^a, a^a,\E^{a}, \H^{a},
 \Sigma_{ab}, \E_{ab}, \H_{ab},\zeta_{ab} \right] \,.
\label{vars} \eea These variables (scalars , 2-vectors and PSTF
2-tensors) form an {\it irreducible set} and completely describe a
vacuum spacetime.\\

In 1+3 formalism, the vector $ u^{a} $ is used to define the
\textit{covariant time derivative} (denoted by a dot) for any tensor
$ T^{a..b}{}_{c..d} $ along the observers' worldlines defined by \be
\dot{T}^{a..b}{}_{c..d}{} = u^{e} \nab_{e} {T}^{a..b}{}_{c..d}~, \ee
and the tensor $ h_{ab} $ is used to define the fully orthogonally
\textit{projected covariant derivative} $D$ for any tensor $
T^{a..b}{}_{c..d} $ , \be D_{e}T^{a..b}{}_{c..d}{} = h^a{}_f
h^p{}_c...h^b{}_g h^q{}_d h^r{}_e \nab_{r} {T}^{f..g}{}_{p..q}~, \ee
with total projection on all the free indices.\\

In 1+1+2 formalism, apart from the `{\it time}' (dot) derivative of an object
(scalar, vector or tensor) which is the derivative along the
timelike congruence $u^a$, we now introduce two new derivatives,
which $ e^{a} $ defines, for any object $ \psi_{a...b}{}^{c...d}  $:
\bea \hat{\psi}_{a..b}{}^{c..d} &\equiv &
e^{f}D_{f}\psi_{a..b}{}^{c..d}~,
\label{hat}\\
\delta_f\psi_{a..b}{}^{c..d} &\equiv &
N_{a}{}^{f}...N_{b}{}^gN_{h}{}^{c}..
N_{i}{}^{d}N_f{}^jD_j\psi_{f..g}{}^{i..j}\;. \eea The hat-derivative
is the derivative along the $e^a$ vector-field in the surfaces
orthogonal to $ u^{a} $. The $\delta$ -derivative is the projected
derivative onto the orthogonal 2-sheet, with the projection on every
free index.

\subsection{Birkhoff Theorem for vacuum LRS-II spacetime}\label{BirkhoffII}
Locally Rotationally Symmetric (LRS) spacetimes \cite{EllisLRS}
exhibit locally (at each point) a unique preferred spatial
direction, which is covariantly defined ~\cite{EllisLRS1}; the
spacetime is invariant under rotations about this direction. Thus
this symmetry requires the vanishing of \textit{all} orthogonal
1+1+2 vectors and tensors, such that there are no preferred
directions in the sheet. Then, all the non-zero 1+1+2 variables are
covariantly defined scalars.

A subclass of the LRS spacetimes, called LRS-II, contains all the
LRS spacetimes that are rotation free; this is the class including
the Schwarzschild solution. As a consequence, in LRS-II spacetimes
the variables $\brac{\A, \Theta,\phi, \Sigma,\E}$ fully characterize
the kinematics (see ~\cite{GosEll11} for the field equations
governing their evolution and propagation). From the field equations
of vacuum LRS-II spacetimes we get a very interesting result: the
1+1+2 scalar of the electric part of the Weyl tensor is always
proportional to (3/2)th power of the Gaussian curvature of the
2-sheet. The proportionality constant sets up a scale in the
problem.\\

To covariantly investigate the geometry of the vacuum LRS-II
spacetime, let us try to solve the {\it Killing equation} for a
Killing vector of the form $\xi_a=\Psi u_a+\Phi e_a$, where $\Psi$
and $\Phi$ are scalars. The Killing equation is
$\nab_{(a}\xi_{b)}=0,$ from which we get the following differential
equations and constraints:

\be \dot\Psi+\A\Phi=0\;\;,\;\; \hat\Psi
-\dot\Phi-\Psi\A+\Phi(\Sigma+\frac13\Theta)= 0,\label{psihat}
\ee

\be \hat\Phi+\Psi(\frac13\Theta+\Sigma)=0\;\;,\;\;
\Psi(\frac23\Theta-\Sigma)+\Phi\phi=0\;.\label{cons2}
\ee

Now we know $\xi_a\xi^a=-\Psi^2+\Phi^2$. If $\xi^a$ is timelike
(that is $\xi_a\xi^a<0$), then because of the arbitrariness in
choosing the vector $u^a$, we can always make $\Phi=0$. On the other
hand, if $\xi^a$ is spacelike (that is  $\xi_a\xi^a>0$), we can make
$\Psi=0$.\\

Let us first assume that $\xi^a$ is timelike and $\Phi=0$. In that
case we know that the solution of equations (\ref{psihat}) always
exists while the constraints (\ref{cons2}) imply that in general,
(for a non trivial $\Psi$), $\Theta=\Sigma=0$. Thus the expansion
and shear of a unit vector field along the timelike Killing vector
direction vanishes. In this case the spacetime is static. Now if $\xi^a$ is
spacelike and $\Psi=0$, we always have a solution for $\Phi$ and
$\phi=\A=0$. Then from the field equations we can deduce that the
spatial derivatives of all quantity vanish and hence the spacetime
is spatially homogeneous. In other words, we can say: {\it There
always exists a Killing vector in the local $[u,e]$ plane for a
vacuum LRS-II spacetime. If the Killing vector is timelike then the
spacetime is locally static, and if the Killing vector is spacelike
the spacetime is locally spatially homogeneous.}\\

If the Gaussian curvature of the sheet is positive, then in the
first case, we can solve the field equations to get a Schwarzschild
exterior ($r>2m)$ vacuum metric, while in the second case we get a
Schwarzschild interior ($r<2m)$ vacuum metric. Thus we have proved
the (local)
\begin{quote}
{\bf Birkhoff Theorem}: {\it Any $C^2$ solution of Einstein's
equations in empty space which is spherically symmetric in an open
set ${\mathcal{S}}$ is locally equivalent to part of maximally
extended Schwarzschild solution in ${\mathcal{S}}$.}
\end{quote}
 It is
interesting to note that the modulus of the proportionality constant
relating the 1+1+2 scalar of the electric part of the Weyl tensor
and the Gaussian curvature is exactly equal to the Schwarzschild
radius.

\subsection{Almost Spherical Symmetry}
To define the notion of an {\it almost spherically symmetric} spacetime,
let us recall that the two dimensional Riemann curvature tensor of a 2-sheet can be written
in terms of the Gaussian curvature `$K$' as
\be
{}^{(2)}R^a_{bcd}=K\left(N^a_cN_{bd}-N^a_dN_{bc}\right)\;.
\label{2riemann}
\ee
Using the above equation we can immediately write the
{\em geodesic deviation equation} for a family of closely spaced
geodesics on the 2-sheets with tangent vectors $\psi^a(v)$ and separation vectors
$\eta^a(v)$ (where `$v$' is the parameter which labels the different geodesics)
as
\be
\psi^e\delta_e(\psi^f\delta_f \eta^a)=K(\psi^a\psi_d\eta^d-\eta^a\psi_c\psi^c)\;.
\label{GDE}
\ee
Let us now define a vector 
\begin{equation}\label{za}
Z^a=\psi^e\delta_e(\psi^f\delta_f \eta^a)-K_0(\psi^a\psi_d\eta^d-\eta^a\psi_c\psi^c).
\end{equation}
 Here $K_0$ is the
Gaussian curvature for a spherical sheet (which is constant for a given value of affine parameter along the integral curves of the vector $e^a$).
If the 2-sheets are exactly spherical then this vector vanishes and hence the magnitude of $Z^a$ (=$\sqrt{Z_aZ^a}$) gives a covariant measure of the
deviation from the spherical symmetry.\\

We now define an {\it almost spherically symmetric}
spacetime in the following way:
\begin{quote}
{\it Any $C^2$ spacetime with positive Gaussian curvature everywhere, which admits a local 1+1+2 splitting at
every point is called an {\underline{almost spherically symmetric}}
spacetime, iff the following quantities are either zero or much smaller than the scale
defined by the modulus of the proportionality constant (that relates
the 1+1+2 Weyl scalar and the (3/2)th power of the Gaussian
curvature).
\begin{itemize}
\item The magnitude of all the 2-vectors (defined by $\sqrt{\psi_a\psi^a}$) and PSTF 2-tensors (defined by $\sqrt{\psi_{ab}\psi^{ab}}$)
described in equation (\ref{vars}).
\item The magnitude of the vector $Z^a$ defined above in (\ref{za}).
\end{itemize}}
 \end{quote}

We would like to emphasize here that though Minkowski spacetime
belongs to the set of LRS-II, in the above definition of the
perturbed spacetime we exclude the Minkowski background, as in that
case the scale is identically zero.

\subsection{Almost Birkhoff theorem for almost spherical symmetry}
The set of all 1+1+2 variables in (\ref{vars}) apart from $\brac{\A,
\Theta,\phi, \Sigma,\E}$ are all of ${\mathcal O}(\epsilon)$ with
respect to the invariant scale. Using  equations (48-81) of
~\cite{CLa07}, we can get the propagation and evolution equations of
these small quantities.\\

In these equations all the zeroth order quantities are background
quantities. If the background is static with $\Theta=\Sigma=0$ and
the time derivative all the background quantities are zero, the time
derivatives of the first order quantities at a given point is of the
same order of smallness as themselves. Hence the first order
quantities still remains  ``small'' as the time evolves. Similarly
if the background is spatially homogeneous with $\phi=\A=0$ and the
spatial derivative all the background quantities are zero, the
spatial derivatives of the first order quantities at a given point
are of the same order of smallness as themselves. Hence the first
order quantities still remain  ``small'' along the spatial
direction. In both the cases of a static background and a spatially
homogeneous background the resultant set of equations are the
perturbed LRS-II equations.\\

Again trying to solve the Killing equation for a Killing
vector of the form $\xi_a=\Psi u_a+\Phi e_a$, we get the following
extra differential equations and constraints (apart from
(\ref{psihat}) and (\ref{cons2})): \be
-\delta_c\Psi+\Psi\A_c+\Phi\brac{\veps_{cd}\Omega^d+\alpha_c+\Sigma_c}=0\;,\label{delpsi}
\ee \be \delta_c\Phi+\Phi a_c +2\Psi\Sigma_c=0\;\;,\;\;
\Psi\Sigma_{cd}+\Phi\zeta_{cd}=0\;.\label{cons3} \ee Now we see that
for both timelike ($\Phi=0$)  or spacelike ($\Psi=0$) vectors, all
the above equations are not completely solved in general for the
arbitrary small values of the first order quantities. However as we
proved that these first order quantities generically remain
${\mathcal O}(\epsilon)$ both in space and time, we can see that a
timelike vector with ($\Theta=\Sigma=0$) or a spacelike vector with
($\phi=\A=0$) almost solves the Killing equations. Therefore we can
say: {\it For an almost spherically symmetric vacuum spacetime there
always exists a vector in the local $[u,e]$ plane which almost
solves the Killing equations. If this vector is timelike then the
spacetime is locally almost static, and if the Killing vector is
spacelike the spacetime is locally almost spatially homogeneous.}\\

Also as we have seen that in this case the resultant set of
equations are the perturbed LRS-II equations, with ${\mathcal
O}(\epsilon)$ terms added to each, and the perturbations locally
remain small both in space and time, a part of the maximally
extended almost-Schwarzschild solution will then solve the field
equations locally. Thus we have proved the (local)
\begin{quote}
{\bf Almost Birkhoff Theorem}: {\it Any $C^2$ solution of Einstein's
vacuum equations 
which is almost spherically
symmetric in an open set ${\mathcal{S}}$, is locally almost
equivalent to part of a maximally extended Schwarzschild solution in
${\mathcal{S}}$.}
\end{quote}

Note that we do not consider perturbations across the horizon: our
result holds for any open set ${\mathcal{S}}$ that does not
intersect the horizon in the background spacetime. The result almost
certainly holds true across the horizon also, but that case needs
separate consideration. The above result can be immediately
generalized in the presence of a cosmological constant. In that case
an `almost' spherically symmetric solution in an open set
${\mathcal{S}}$, is locally almost equivalent to part of a maximally
extended Schwarzschild deSitter/anti-deSitter solution in
${\mathcal{S}}$. Thus the 2-mass solution above is stable to
inhomogeneous perturbations in this sense.\\

What are the implications? First, the role of Birkhoff's Theorem in
astrophysics, characterizing the gravitational field in the vicinity
of massive objects, is unchanged due to geometric perturbations: for
example those due to Jupiter in the vicinity of the Sun or stellar rotation, which
has considerable effects in perturbing spherical symmetry near the vicinity
of the stars. In other words, the rigidity of spherical vacuum manifolds in General relativity
continues even in the perturbed scenario.
Secondly,
this result is unlikely to play any role as regards the expansion of
the universe : such  perturbations probably don't affect the way
local static domains add up to give an expanding universe (for example rotational effects are stationary and do not help in explaining expansion).

\section{Birkhoff with matter: finite infinity}\label{sec:finitge infinty} %

We know that real astronomical systems are neither exactly spherically
symmetric, nor exactly empty. While the Birkhoff theorem and its generalization remains valid for the case of
an elecrovac solution (\cite{Din92}, section 18.1), Birkhoff's
theorem is not true in general when matter is present, as is shown
for example by the Lema\^{i}tre-Tolman-Bondi solutions \cite{LTB,Kra97}.
It remains true if the matter is static (\cite{CapFar11}, Section
4.3) but this will not be true in general. These results do not
include crucial cases such as the Solar System, which is neither
exactly empty nor exactly spherically symmetric.\\

In a previous section we showed that the result is stable to small
geometric perturbations: it remains true if spacetime is not exactly
spherically symmetric. Here we summarize \cite{GosEll12}, which
shows that the result is stable to small matter perturbations: it
remains true if spacetime is not exactly vacuum, as for example in
the case of the solar system.\\

In other words, we would like to ask the question, how much matter
can be present if the Birkhoff theorem is to remain approximately
true. That is, we would like to perturb a vacuum LRS-II spacetime by
introducing a small amount of general matter in the spacetime. In
this section we only deal with the static exterior background as
that is astrophysically more interesting case.

\subsection{Matter}

In the 1+3 splitting, the Energy Momentum Tensor $T_{ab}$ of a
general matter field can be written as \be T_{ab}=\mu
u_au_b+q_au_b+u_aq_b+p h_{ab}+\pi_{ab} \ee Where the scalars
$\mu=T_{ab}u^au^b$ and $p=(1/3)T_{ab}h^{ab}$ are the energy density
and isotropic pressure respectively. The 3-vector,
$q^a=T_{cb}u^bh^{ca}$, is the heat flux and the PSTF 3-tensor,
$\pi_{ab}=T_{cd}h^c_{<a}h^d_{b>}$, defines the anisotropic stress.
In 1+1+2 splitting of LRS-II spacetime, we can write the fluid
variables as $q^a=Qe^a$ and $\pi_{ab}=\Pi\bras{ e_{a}e_{b} -\frac12
N_{ab}} $, where $Q$ and $\Pi$ are  scalars.\\

 We know from the covariant linear perturbation
theory, any quantity which is zero in the background is considered
as a first order quantity and is automatically gauge-invariant by
virtue of the Stewart and Walker lemma \cite{SW}. Hence the set
$\brac{\Theta,\Sigma,\mu, p, \Pi, Q }$, describes the first order
quantities, on a vacuum LRS-II background. As we have already seen,
the vacuum spacetime has a covariant scale given by the
Schwarzschild radius which sets up the scale for perturbation. Let
us locally introduce general matter on a static Schwarzschild
background such that \be \left[ \frac{\mu}{K^{(3/2)}},
\frac{|p|}{K^{(3/2)}}, \frac{|\Pi|}{K^{(3/2)}},
\frac{|Q|}{K^{(3/2)}}\right]<< C, \label{cond1} \ee and \be \left[
\frac{|\hat\mu|}{K^{(3/2)}}, \frac{|\hat p|}{K^{(3/2)}},
\frac{|\hat\Pi|}{K^{(3/2)}} \frac{|\hat Q|}{K^{(3/2)}}, \right]<<
\phi C \label{cond2} \ee where $C$ is the proportionality constant
described in the previous section (that relates
the 1+1+2 Weyl scalar and the (3/2)th power of the local Gaussian
curvature), which is also the Schwarzschild
radius, and $K$ is the local Gaussian curvature.

\subsection{Domains}
Now we need to make clear in what domain these equations will hold.
The application will be to the spherically symmetric exterior domain
of a star of mass $M$ and Schwarzschild radius $R_S = 2M$, in the
units of $8\pi G=c=1$. We will define \emph{Finite Infinity} ${\cal
F}$ as a 2-sphere of radius $R_{\cal F} \gg R_M$ surrounding the
star: this is infinity for all practical\ purposes
\cite{Ell84,StoEll}. We assume the relations (\ref{cond1},
\ref{cond2}) hold in the domain $D_{\cal F}$ defined by $r_S < r <
R_{\cal F}$ where $r_S > r_M $ is the radius of the surface of the
star. This is the local domain where our results will apply. In the
case of the solar system, $R_{\cal F}$
can be taken to be about a light year .\\

It is important to make this restriction, else eventually we will
reach a radius $r$ where these inequalities will no longer hold;
in the real universe asymptotically flat
regions are always of finite size, being replaced at larger scales
by galactic and cosmological conditions. The result we wish to prove
is a local result, applicable to the locally restricted nature of
real physical systems.

\subsection{Matter perturbations on vacuum LRS-II spacetimes}

From the linearised matter conservation equations of the system  (
see ~\cite{GosEll12} for detailed description of these equations) we
can see that if (\ref{cond1}) and (\ref{cond2}) are locally
satisfied at any epoch, within the domain $D_{\cal F}$, then the
time variation of the matter variables are of same order of
smallness as themselves. Hence there exists an open set
$\mathcal{S}$ within where the amount of matter remains ``{\it
small}'', if the amount is small at any epoch in $\mathcal{S}$ and
only small amounts of matter enter $D_{\cal F}$ across ${\cal F}$.\\

One could attempt to determine the same kinds of inequality as those
above for matter crossing ${\cal F}$, but one can resolve this issue
in another way: we have not yet specified the time evolution of
${\cal F}$. We now do so in the following manner: choose it in a
suitable manner in some initial surface $t = t_0$, and then
propagate it to the future by dragging it along world lines that are
integral curves of the timelike eigenvector of the Ricci tensor
$R_{ab}$ (this will be unique for any realistic non-zero matter). As
these are then timelike eigenvectors of the stress tensor $T_{ab}$
(because of the field equations), equal amount of energy density
will convect in and out across ${\cal F}$ due to random motions of
matter \cite{Covariant}; the total amount of matter inside ${\cal
F}$ will be conserved, and if the inequalities (\ref{cond1},
\ref{cond2}) are satisfied at some initial time they will be
satisfied at later times, unless major masses enter the ${\cal F}$
locally in some region. If this is so,
we do not have an isolated system and the extended Birkhoff's theorem need not apply.\\

Hence we will define the time evolution of ${\cal F}$ in the way
just indicated, and suppose that (\ref{cond1}, \ref{cond2}) are then
satisfied at later times; if this is not the case the local system
considered is not isolated and our result is not applicable.

\subsection{Almost symmetries}
From the linearised Field equations, it is evident that if the
matter variables remain ``{\it small}'' as defined by (\ref{cond1})
then the spatial and temporal variance of the expansion $\Theta$ and
the shear $\Sigma$ are of the same order of smallness as the matter.
In that case we see that a timelike vector will not exactly solve
the Killing equations (\ref{psihat})-(\ref{cons2}) in general,
although it may do so approximately.\\

To see this explicitly, let us set $\Phi = 0$ in Killing's equation 
and consider the following symmetric tensor $K_{ab} := \nab_a(\Psi
u_b) + \nab_b(\Psi u_a) $. This tensor vanishes if $\Psi u^a$ is a
Killing vector. This is the case of an exact symmetry when the
spacetime is exactly static. However, in the perturbed scenario, to
see how close the vector $\xi_a = \Psi u_a$ is to a Killing vector,
let us consider the scalars constructed by contracting the above
tensor by the vectors $u^a$, $e^a$ and the projection tensor
$N^{ab}$. If the conditions \be
\left[\frac{|K_{ab}u^au^b|^2}{K^{3/2}},\frac{|K_{ab}u^ae^b|^2}{K^{3/2}},
\frac{|K_{ab}e^ae^b|^2}{K^{3/2}},
\frac{|K_{ab}N^{ab}|^2}{K^{3/2}}\right]<<C \label{cond3} \ee are
satisfied, then we can say that $\xi_a = \Psi u_a$ is close to a
Killing vector and the spacetime is approximately static. From the
previous section (using equation (\ref{psihat})) we know that there
always exist a non-trivial solution of the scalar $\Psi$ for which
$|K_{ab}u^au^b|$ and $|K_{ab}u^ae^b|$ vanishes; we choose $\Psi$
accordingly. However for a general matter perturbation, as $\Theta$
and $\Sigma$ are non-zero,  $|K_{ab}e^ae^b|^2$ and
$|K_{ab}N^{ab}|^2$ are generally non-zero. However, using the
linearised field equations we get \be \left(\frac13 \Theta - \frac12
\Sigma\right)^{2}~\approx  \frac13 \mu- \frac12 \Pi.\;;\;
\left(\frac13\Theta+\Sigma\right)\left(\frac23\Theta-\Sigma\right)~\approx
\frac23\mu +\frac12\Pi\;. \label{KE12} \ee
 Thus we see that if the amount
of matter is ``small'', that is the condition (\ref{cond1}) is
satisfied, then the following conditions are satisfied \be
|K_{ab}e^ae^b|^2=\Psi^2(\frac13\Theta+\Sigma)^2\ll C K^{3/2},
\label{cons12} \ee \be
|K_{ab}N^{ab}|^2=\Psi^2(\frac23\Theta-\Sigma)^2\ll C K^{3/2}.
\label{cons22} \ee Therefore we can say that there always exists a
timelike vector that satisfies (\ref{cond3}). This vector then
almost solves the Killing equations in $\mathcal{S}$ and hence the
spacetime is {\it almost} static in $\mathcal{S}$. This is the
\textbf{Almost Birkhoff theorem} for an almost vacuum spherically
symmetric
solution.\\

The above conditions, (\ref{cond1}) and (\ref{cond2}), can also be
written in another way. \be \left[ \frac{|R|}{K^{(3/2)}},
\frac{|R_{ab}u^au^b|}{K^{(3/2)}},
\frac{|R_{<ab>}e^ae^b|}{K^{(3/2)}},
\frac{|R_{<ab>}u^ae^b|}{K^{(3/2)}} \right]<< C \label{cond4} \ee and
\be \left[ \frac{|\hat
R|}{K^{(3/2)}},\frac{|R_{ab}u^au^b|\hat{}}{K^{(3/2)}},
\frac{|R_{<ab>}u^ae^b|\hat{}}{K^{(3/2)}},\frac{|R_{<ab>}e^ae^b|\hat{}}
{K^{(3/2)}}\right]<< \phi C \label{cond5} \ee In other words the
ratio of the scalars constructed from the Ricci tensor using the
vectors $u^a$ and $e^a$ (and their spatial variations) to the
$(3/2)$th power of the local Gaussian curvature of the 2-sheet
should be much smaller than the Schwarzschild radius if the Birkhoff
theorem is to remain approximately true. Equations (\ref{cond4}) and
(\ref{cond5}) are easier to use, in case of presence of multifluids
in the spacetime.\\

What are the implications? First, the role of Birkhoff's Theorem in
astrophysics, characterizing the gravitational field in the vicinity
of massive objects, is unchanged due to small matter perturbations,
for example dust or dark matter pervading the solar system.
Secondly, the solution is almost but not exactly static, and this might indeed play a  role as regards the expansion
of the universe. Dark matter could conceivably affect the local
static domains so they each give a small contribution to an expanding
universe, which adds up to give the global effect we see. But then
we should be able to measure the Hubble constant in the solar
system, at least in principle (See Section \ref{darkmatter}),

\subsection{Comments on the solar system}

In case of the solar system  ~\cite{SS} we know that the
interplanetary medium includes interplanetary dust, cosmic rays and
hot plasma from the solar wind. Its density is very low at about 5
particles per cubic centimeter in the vicinity of the Earth; it
decreases with increasing distance from the sun, in inverse
proportion to the square of the distance. In this section, to
compare our result with the observed astronomical data, we will use
SI units for clarity.\\

The density of interplanetary medium is variable, and may be
affected by magnetic fields and events such as coronal mass
ejections. It may rise to as high as 100 particles/$cm^3$. These
particles are mostly Hydrogen nuclei, and hence the maximum density
per cubic meter will be approximately of the order of $10^{-19}$
Kilograms, and the local Gaussian curvature of the heliocentric
celestial sphere in the vicinity of the earth is of the order of
$10^{-22}\;\;m^{-2}$ . Hence the ratio of the maximum interplanetary
density to the  $(3/2)$th power of the local Gaussian curvature is
of the order of $10^{14}$ Kilograms, which is much smaller then the
solar mass ($10^{30}$ Kilograms). Also the large amplitude waves in
the medium are comparable to the energy density of the unperturbed
medium, which makes the spatial variation of energy density to be of
the same order of smallness as itself. This satisfies (\ref{cond1})
and (\ref{cond2}) and hence in the solar
system the Birkhoff theorem remains almost true.\\

We can relate { the discussion} to the \emph{Finite Infinity}
concept { for the solar system}. We know that the outer edge of
the solar system is the boundary between the flow of the solar wind
and the diffused interstellar medium. This boundary, which is known
as the {\it Heliopause}, is at a radius of approximately $10^{13}$
meters. The interplanetary medium thus fills the roughly spherical
volume contained within the heliopause. As the density of the
interplanetary medium decreases in inverse proportion to the square
of the distance, the density near the heliopause is of the order of
$10^{-23}$ Kilograms per cubic meter.  Hence the ratio of the
density to the  $(3/2)$th power of the local Gaussian curvature is
of the order of $10^{16}$ Kilograms and still remains much smaller
than the solar mass. Also the amount of matter crossing the
heliopause to the diffused interstellar medium is of the same order.
Hence we can easily define the heliopause as the boundary of our
domain $D_{\mathcal{F}}$. As the conditions
 (\ref{cond1}) and (\ref{cond2}) { are} 
 true at the boundary of the domain,
they should be true everywhere inside the domain, unless the matter
outside the star is highly clustered locally. But we are considering
the case 
{of} a low density diffuse gas where this is not the
case. {the conditions (\ref{cond4}) and (\ref{cond5}) will be satisfied in this domain.}\\

For the massive planets inside the solar system (e.g. Jupiter or
Saturn), these conditions may be violated in their very close
vicinity, but in that case the local spacetime no longer remains
spherically symmetric. This can be easily checked by calculating the geodesic deviation equation
near the Lagrangian points of these planets and calculating the magnitude of the vector $Z^a$
described in the previous section.
 However as the vast fraction of the solar
system's mass (more than $99\%$) is in the sun, on average these
massive planets have a very tiny effect on the system as a whole and
the approximate theorem remains true. Hence the local spacetime
within the solar system is ``{\it almost}'' described by a
Schwarschild metric. Note that we have not included dark matter in
this mass budget. We are unaware of any claims that it is
significant in the solar system.
\section{The expansion of the universe}\label{sec:summary} %
Birkhoff's theorem is a key theorem relating the global expansion of
the universe to local gravitating systems: it protects them from the
expansion of the universe, and raises interesting issues as to how
such static domains can be patched together to give an expanding
universe.\\

We have explored these relations here to see how they can be
compatible, inter alia summarising two useful generalisations of
Birkhoff's theorem. The point that arises is whether dark matter -
the dominant form of matter in the universe at large scales -
pervades the solar system and links our local system to the cosmic
expansion. This seems unlikely -- but if it is true, cosmology can
in principle be done in the solar system by measuring its
time-changing effects on solar system dynamics.\\

\textbf{Acknowledgements:} We thank Chris Clarkson for helpful comments, and the NRF and UCT Research Committee for financial support.


gfre::version \emph{2013-04-10}
\end{document}